\expandafter\ifx\csname natexlab\endcsname\relax\fi

\documentclass[12pt,preprint]{aastex}

\def\Msol{\thinspace\hbox{$\hbox{M}_{\odot}$}}

\def\a4{\hsize 17.0cm \vsize 25.cm}

\shorttitle{Supernovae blast waves in young globular clusters}
\shortauthors{Tenorio-Tagle  et al.}

\begin{document}

\title{Supernovae and their expanding blast waves during the early
       evolution of Galactic globular clusters}

\author{Guillermo Tenorio-Tagle\altaffilmark{1}, Casiana 
Mu\~noz-Tu\~n\'on\altaffilmark{2}, Sergiy Silich\altaffilmark{1},  
Santi Cassisi\altaffilmark{3}}

\affil{$^1$Instituto Nacional de Astrof\'\i sica \'Optica y
Electr\'onica, AP 51, 72000 Puebla, M\'exico; gtt@inaoep.mx}
\affil{$^2$Instituto de Astrof\'\i sica de Canarias
cmt@iac.es}
\affil{$^3$INAF – Astronomical Observatory of Collurania, via 
M. Maggini, 64100 Teramo, Italy; cassisi@oa-teramo.inaf.it}

\begin{abstract}
Our arguments deal with the early evolution of 
Galactic globular clusters and  show why only a few of the 
supernovae products were retained within globular clusters and only in the 
most massive cases ($M \ge 10^6$ M$_\odot$), while less massive clusters were 
not contaminated at all by supernovae. Here we show that supernova blast waves
evolving in a steep density gradient undergo blowout and end up discharging 
their energy and metals into the medium surrounding the clusters. This 
inhibits the dispersal and the contamination of the gas left over from a 
first stellar generation. Only the ejecta from well centered supernovae, that
evolve into a high density medium available for a second stellar generation in
the most  massive clusters would be retained. These are likely to mix their 
products with the remaining gas, leading in these cases 
eventually to an Fe contaminated second stellar generation. 
\end{abstract}


\section{Introduction}
\label{sec:1}

In the last decade, a huge observational evidence has deeply challenged the 
common paradigm of Galactic Globular Clusters (GGCs) as the prototype of 
Single Stellar  Populations. High resolution spectroscopy \citep[see][for a 
review]{Gratton2012}
and high accuracy photometry \citep[see][and references therein]{Piotto2012}
have actually shown that GGCs host samples of stars with 
distinct light-element chemical patterns, populating distinct evolutionary 
sequences evident  in UV-optical Color-Magnitude diagrams 
\citet[][]{Sbordone2011}; \citet[][]{Piotto2015}.
In the bulk of GGCs, the distinct
sub-populations in any given cluster - characterised by their peculiar 
chemical patterns - do not show any difference in their overall metallicity as
commonly traced by the Fe abundance and thus present no evidence of 
contamination by supernovae (SNe). The implication is that the homogeneously 
low [Fe/H] value in all their stars is probably original to their 
proto-cluster primordial cloud. Note however that given the mass of GGCs, 
ranging  from a few times 10$^5$ M$_\odot$ to several times 10$^6$ M$_\odot$ 
and any reasonable initial mass function (IMF), the number of 
type II SNe expected during their  early 40 Myr of evolution, amounts 
from a few thousands to a few tens of thousands of events and even so, non
of their products appear to have been trapped by the sub-populations. A large 
intrinsic difference in iron abundance (${\rm [Fe/H] \ge 0.1} \sim$ dex) where
SNe have played a major role in the chemical enrichment 
history of the cluster,  has been considered for a long time as a peculiarity 
of the most massive GGC $\omega$ Cen \citep[see][and references 
therein]{Marino2011}.
This is now changing given the evidence of other massive GGCs with 
internal variations in their metallicity. Such is the case 
of M54 \citep[][]{Carretta2010}, Terzan 5 \citep[][]{Ferraro2009}, 
M2 \citep[][]{Yong2014}, NGC 5824 \citep[][]{DaCosta2014}, M22 
\citep[][]{Marino2009} and NGC 5286 \citep[][]{Marino2015}.
A scenario(s) able to explain the events leading  to star clusters hosting 
multiple stellar populations (with their intriguing photometric and 
spectroscopic peculiarities) is still a matter of debate (see 
\citet[][]{Cassisi2013}, \citet[][]{Renzini2013} and references therein).
For instance, to understand why only some GCs were able to retain the ejecta
of the SNe associated to a first stellar generation (1SG) and thus  increase 
the metallicity of the following second (and in some cases also third) 
generation (2SG) is obviously of pivotal importance. At the same time, as 
discussed by \citep[][]{Renzini2013},
the observed metallicity spread in clusters with \lq{multiple [Fe/H] 
abundances}\rq\ poses a stringent constraint on the efficiency with which 
these GCs were able to retained the SN ejecta: the fraction of the ejecta 
expelled by the whole population of SNe and retained by the most massive GCs 
is only of the order of a few percent\footnote{In the case of 
$\omega$ Cen - one of the clusters with the largest metallicity spread, 
\citep[][]{Renzini2013}
estimated that  only $\sim 0.2$\% of the ejecta of Type II SNe belonging to 
the 1SG has to be retained in order to justify the observed metallicity 
distribution.}.

Here we focus on issues that could provide important clues about the 
properties of star clusters during their early stages. The aim is to explain 
why only some GGCs retain a small fraction of the SN ejecta while the bulk of 
GGCs did not retain them at all. For this we explore the hydrodynamics of SN 
blast waves as they sweep the surrounding medium and  look for the conditions 
that may inhibit their usually expected  dispersing nature.

\section{Blowout and the evolution of Superbubbles}
\label{sec:2}

An explosion in an inhomogeneous atmosphere could lead to the acceleration of 
the leading shock into the density gradient.  
Numerical simulations \citep[see][and references therein]{Tenorio1988}
have considered an exponential or a gaussian galactic HI disk 
surrounded by a large gaseous halo and a massive young stellar cluster at or 
near the galaxy plane, driving a supersonic wind. The latter first generates 
a strong shock into the surrounding medium that steady decelerates, in all 
directions, as it continuously sweeps more of the surrounding gas into an 
expanding shell.
However, as shown by \citet[][]{Koo1992},
if the shock and its shell reach one scale-height (a couple of hundred pc)  
moving still supersonically, then the shock  accelerates. This is because as 
the violent expansion proceeds, the pressure driving the shock indeed decays, 
but  if the density in the direction away from the galaxy plane falls  
even more rapidly, then the section of the shock moving in that direction, 
instead of decelerating, as in explosions in a constant density medium, it 
would accelerate 
($V_{shock} \propto (P/\rho)^{0.5}$) and the more so, the smaller that 
$\rho$ becomes as the expansion proceeds.
The sudden acceleration initiates blowout: shock acceleration sets a 
Rayleigh-Taylor instability in the leading section of the shell and this 
causes soon its fragmentation. The multiple piercings on the shell allow  for 
the venting, between shell fragments, of the hot ejecta out of the bubble. 
Once this happens the shell stalls. Only a small fraction of the matter swept 
prior to blowout  is accelerated away from the galaxy while most of it ($\sim 
95\%$) remains in the perturbed fragmented shell at the height acquired prior 
to blowout \citep{MacLow1989}.

\section{SN blast waves  evolving in a strong density gradient}
\label{sec:2.}

If the efficiency of star formation in a GC first generation has been about 
50$\%$, then all
the stars including potential core-collapse SNe will evolve 
buried by the large amount of gas left over from star formation. 
Here we assume that this conforms a centrally condensed cloud that spans 
across the cluster  radius $R_{SC}$ and presents a  
gaussian density distribution: $\rho_{cl} = \rho_0 \exp[-(r/R_c)^2]$, 
where $\rho_0$ and $R_c$ are the central gas density and the cluster core 
radius and the whole distribution is surrounded by a low density ambient
medium with $\rho_{ISM} << \rho_0$. The mass of the cloud then is:
\begin{eqnarray}
      \nonumber
      & & 
M_{cl} = 4 \pi \rho_0 \int_0^{R_{SC}} \exp[-(r/R_c)^2] r^2 {\rm d} r =
      \\[0.2cm] & & 
      \label{eq2}
         4 \pi \rho_0 R_c^3 [\pi^{1/2} erf(R_{SC}/R_c)/4 -
         R_{SC}/(2 R_c) \exp[-(R_{SC}/R_c)^2]] , 
\end{eqnarray}
where $erf(R_{SC}/R_c)$ is the error function, e.g. $M_{cl} \approx 2 \times 
10^5$M$_{\odot}$ if one assumes, as in our reference model, the central gas 
number density $n_0 = 10^6$~cm$^{-3}$ and $R_c = 1$~pc.
We also assume that most of the massive stars are binaries 
\citep[as in][]{deMink2009}.
This is expected for a large stellar population restricted to a 
very small volume.
Contrary to single stars, massive binaries deposit their H burning products 
with very low
velocities, favouring mixing with the remaining gas without perturbing its
overall density distribution. Thus we envisage a strongly concentrated gas
density distribution hardly affected by the Roche lobes around massive 
binaries. This allow us to look for the effects produced by single SN. We 
ignored the short phases that follow mass transfer and that likely lead to 
excavated bubbles around exploding stars. Such structures, given the large cloud densities, would be small. However they will 
extend the free-expansion phase of the SN ejecta and
upon the ejecta-swept up shell interaction a sudden thermalisation of the released kinetic energy would take place. Finally, and depending on the mass in 
the surrounding shell, the Sedov phase may be completely avoided, making the 
strongly radiative SN enter its snowplow phase, to then progress 
supersonically into our assumed cloud density distribution 
\citep[see][]{Tenorio1991},
similar to the initial condition in our calculations. Our 
approach differs from  that of \citet{Krause2012,Krause2013}
who consider a 
continuous central wind powered by single massive stars and sequential SNe and 
also by the energy delibered  by accretion into black holes and 
neutron stars. Such an injection of energy led to the build up of 
superbubbles that in all their cases experienced blowout and thus 
led to no contamination of the gas left over from a 1SG, leaving without an 
explanation the Fe spread in the most massive GCs.
   
Here we show that single supernova blast waves evolving in a medium with a 
steep density gradient  
also naturally experience blowout, i. e. the suden and continuous acceleration
 of the 
leading shock in the direction of lowest densities and the fragmentation of 
the leading section of the shell. The high pressure gas, the thermalised 
ejecta, 
rapidly escapes between shell fragments to closely follow the shock into the 
gradient as this becomes  more elongated. Given the size of the cluster 
($R_{SC}$), the ejecta are then expelled into the surrounding medium soon 
after blowout.
This leads to a rapid loss of energy and pressure from the volume swept by the
 SN shock which
strongly inhibits the lateral growth of the cavity.  In fact, soon after 
blowout, the now larger 
pressure of the dense cloud matter surrounding the cavity, would favor its 
collapse towards the elongated volume symmetry axis, leaving then no trace of 
the SN explosion. Explosions occurring within the cluster core however, would 
require that their leading shocks reach  the cluster core boundary 
and enter the density gradient while still progressing with a supersonic 
velocity, for such an evolution to take place. 
In both cases the SN products would be expelled from the cluster.  

Only SN blast waves well contained near the cluster center while evolving 
into the dense 
background medium available for the formation of a 2SG would contribute
to eventually enhance its Fe abundance. This occurs if
the SN blast wave does not reach supersonically the cluster core radius. In 
a dense proto-cluster
cloud environment the swept-up gas cools down and collapses into a 
cold, dense and narrow shell very rapidly, making an early transition from the 
quasi-adiabatic Sedov to the snowplow evolution at $t_0 = 3 k T_s / (4 n_{off} \Lambda(T_s, Z))$, 
where $k$ is the Botzmann constant, $\Lambda(T_s,Z)$ is the cooling function,
$Z$ is the shocked gas metallicity, $n_{off}$ is the gas number density at the
explosion site and $T_s$ is the post-shock temperature: $T_s = 2(\gamma-1)  \eta V_s^2 / ((\gamma+1)^2 k)$.
Here $\eta = 14/23 m_H$ is the mean mass per particle in the shocked ionized 
plasma with 10 hydrogen atoms per each helium atom, $m_H$ is the proton mass,
$\gamma = 5/3$ is the ratio of specific heats and $V_s = 0.4 
(\xi E_0 / \rho_c)^{1/5} t_0^{-3/5}$ is the velocity of the SN 
blast wave. Inside the cloud core radius $R_c$, where the density 
distribution is almost homogeneous, the velocity of the swept-up shell is 
\citep[e.g.][]{Pasko1986}:
\begin{equation}
\label{eq9}
V_s = \frac{2}{7}\left(\frac{147 E_{T0}R^2_0}{4\pi\rho_c}\right)^{1/7} 
      t^{-5/7} ,
\end{equation}
where $R_0 = (\xi E_0 / \rho_c)^{1/5} t_0^{2/5}$ and 
$E_{T0} = (\gamma+1)E_0 / (3\gamma-1)$ are the radius of the shock and the 
thermal energy driving the expansion  
at the end of the Sedov phase,
$E_0 = 10^{51}$~erg is the energy of the explosion, $\rho_c$ is the central 
gas density in the proto-cluster cloud and
$\xi = 75(\gamma-1)(\gamma+1)^2/(16\pi(3\gamma-1))$ 
\citep[e.g.][]{Bisnovatyi1995}.
The velocity drops to the sound speed value $a_s$ and the shell stalls at
\begin{equation}
\label{eq10}
R_{stall} = \left(\frac{2}{7a_s}\right)^{2/5}
             \left(\frac{147 E_{T0}R^2_0}{4\pi\rho_c}\right)^{1/5} .
\end{equation}
The stalling radius $R_{stall}$ is smaller than the core radius $R_c$  if 
the central density exceeds the critical value:
\begin{equation}
\label{eq11}
\rho_{crit} = \left(\frac{2}{7a_s}\right)^{10/7}
             \left(\frac{147 E_{T0}}{4\pi}\right)^{5/7} 
             \left(\frac{\xi^2 E^2_0 t^4_0}{R^{25}_c}\right)^{1/7}  ,
\end{equation}
Equation (\ref{eq11}) allows one to calculate the critical mass of the cloud
$M_{crit}$. Only globular clusters with a gas mass available for a second 
stellar generation  $M > M_{crit}$ would be able to retain SN products  and 
eventually enhance its Fe abundance. This may help to explain the observed Fe 
spread at the high-mass end of GGCs and the lack of it for the bulk of GCs.  
For a gas temperature $T_{cloud}=100$~K and a cloud core radius between
0.5pc $\le R_c \le$ 2pc, equation (\ref{eq11}) leads to a critical mass
$(2-5) \times 10^5$\Msol.

To calculate the SN shock evolution for off centered explosions, we used our 
2D thin layer approximation code \citep{Bisnovatyi1995}.
 In all 
cases  the transition to the snow-plow phase occurs when the shock radius is 
still small (0.045~pc $\le R_{cool} \le 0.3$~pc) and thus for the calculations
we adopted as initial conditions the shock radius and expansion velocity at 
the end of the Sedov evolution evaluated for the gas density  $\rho_{off}$ at 
the SN site.

Figure 1 shows our results for a proto-cluster 
cloud with a central gas density $n_0 = 10^6$~cm$^{-3}$ and $R_c = 1$pc,
with a SN explosion first at the center of the gas distribution and 
then at 1 pc and at 2 pc away from the 
center.
The Figure tracks the calculated shape of the SN blast wave at three different 
times (solid, dotted and dashed lines) as well as the spherically symmetric 
gaussian density distribution of the gas cloud, shown
by  thin concentric lines labeled with the logarithmic values of the local
density. Also shown are the top pole shock velocity as a function of distance 
to the GC center. The latter plots display the strong shock acceleration and 
thus blowout,  as the shock enters the steep density distribution, whereas
in the first case, the shock velocity goes to 0 km s$^{-1}$ 
(see Figure 1b) inside $R_c$ and thus the SN products are trapped to finally
contaminate the remaining cloud. Other calculations assuming the explosion
site to be $\le$ 0.25 pc have lead to the same result, well contained SN 
explosions, while explosions out of this radius sooner or later experience 
blowout. Comparing this volume with that of the cloud, the proportion 
of expected well-contained SN is only 0.024\%. More massive clouds with a 
larger trapping radius will lead to a larger proportion of SN contaminating 
the cloud, as expected for the most massive GGCs.   
\begin{figure}[htbp]
\vspace{18.0cm}
\includegraphics{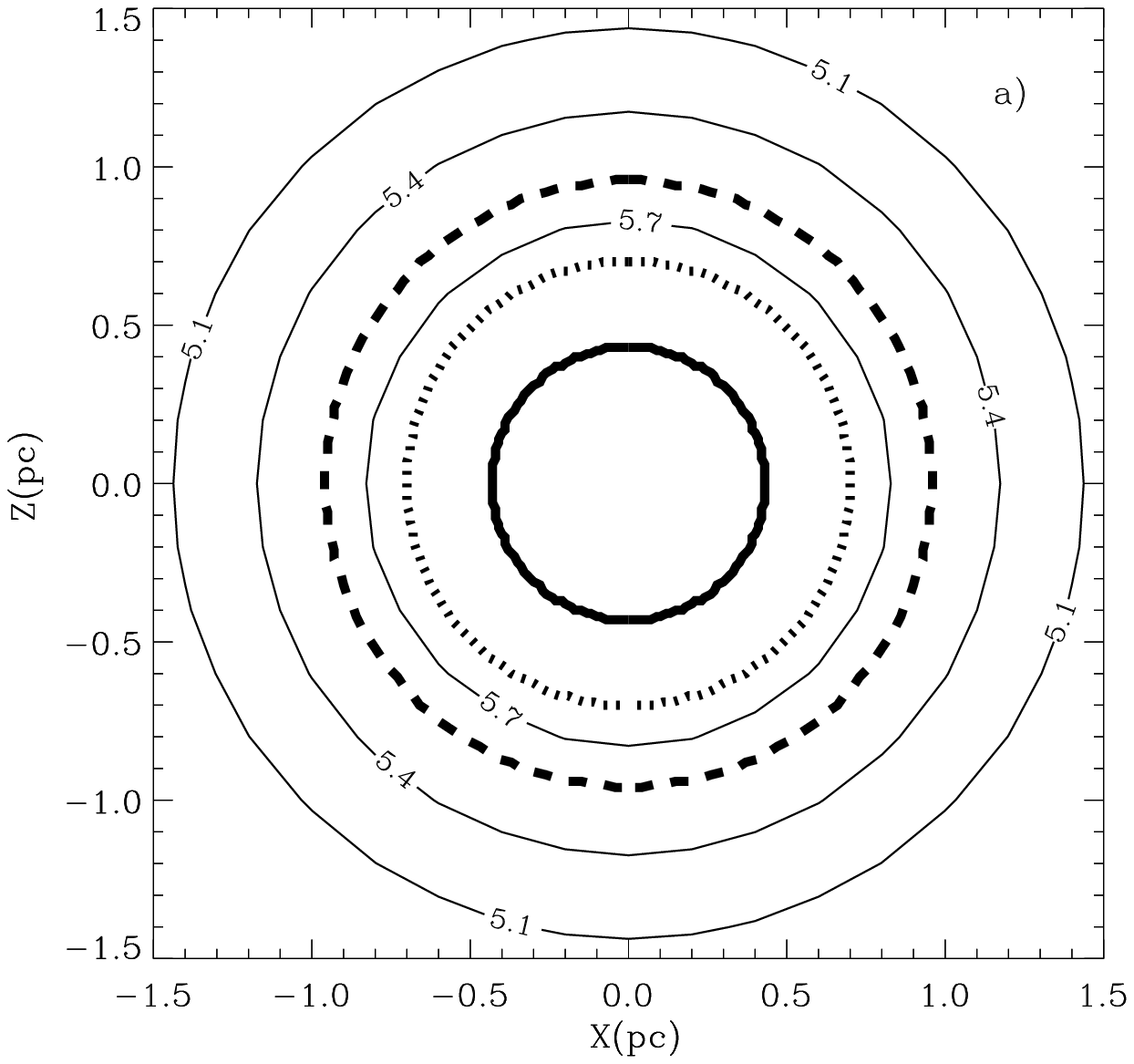}
\includegraphics{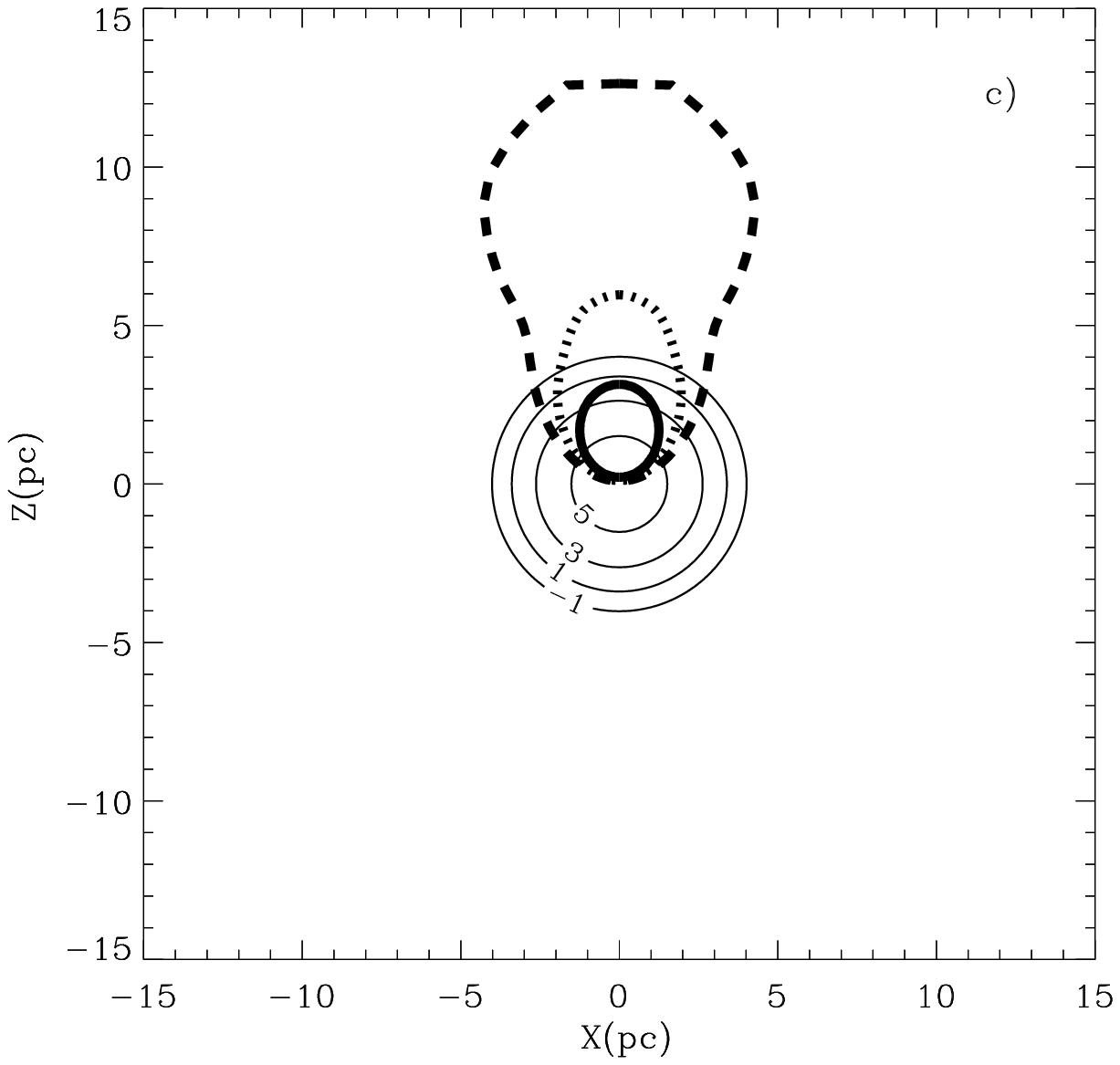}
\includegraphics{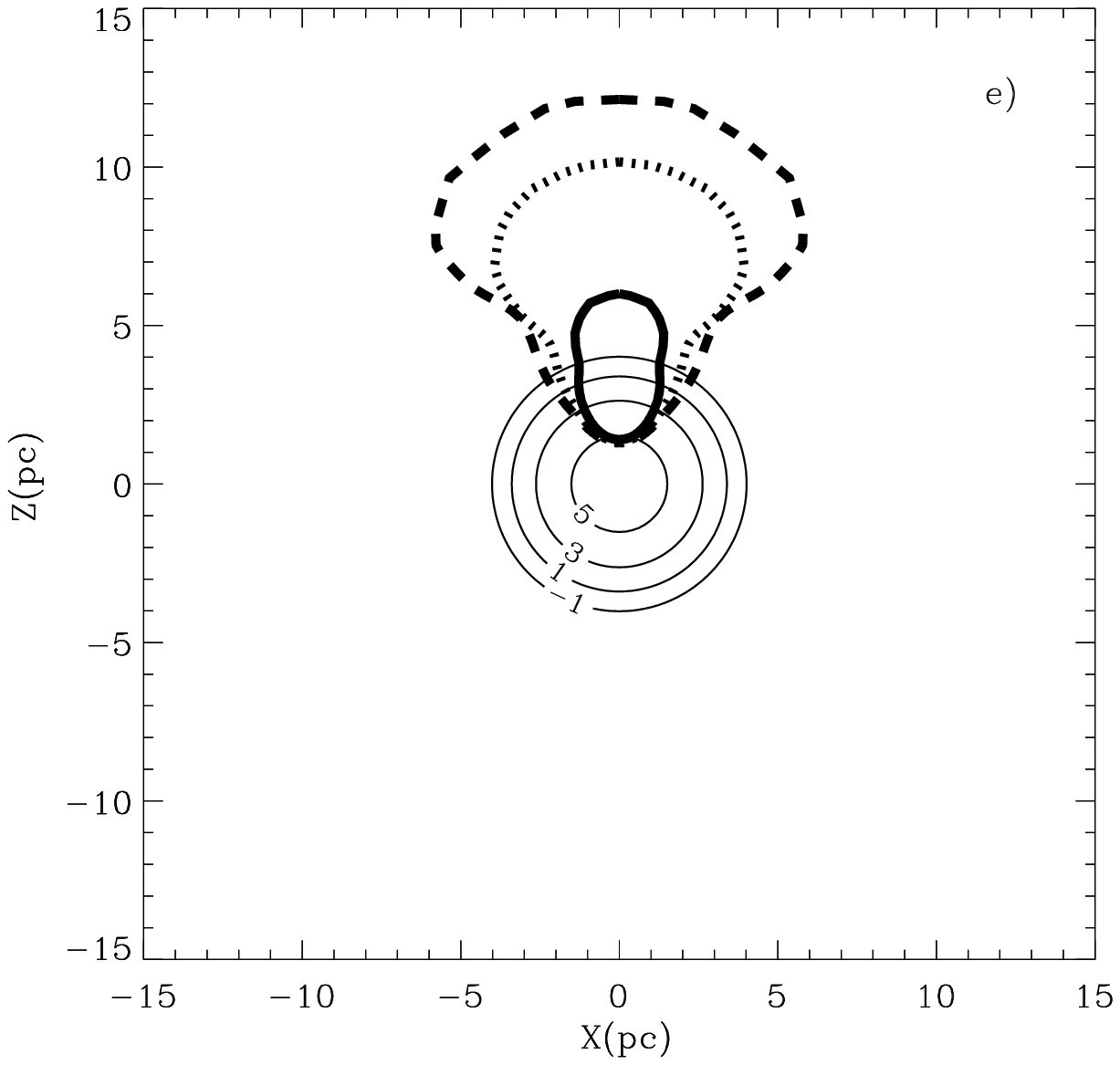}
\includegraphics{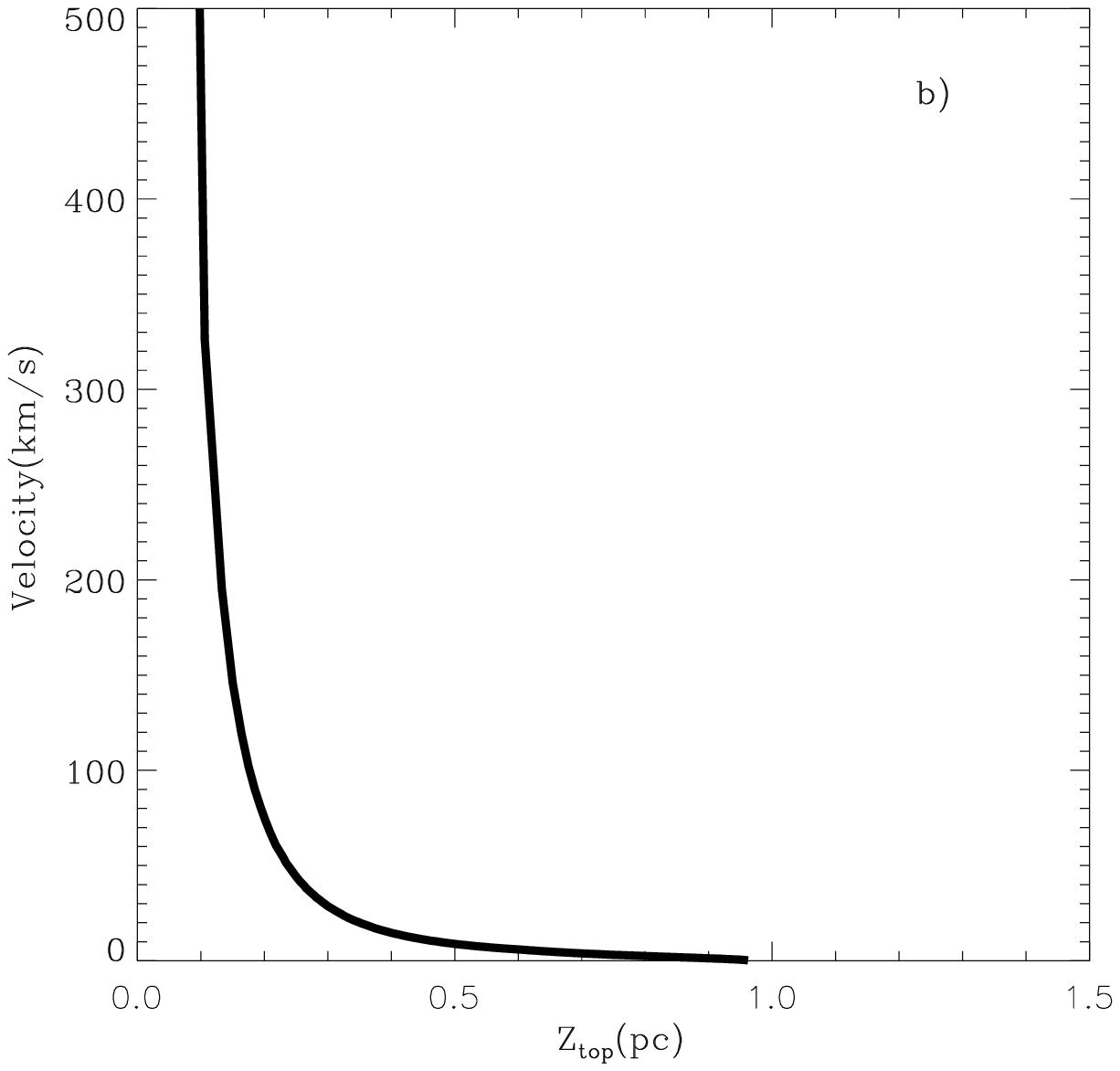}
\includegraphics{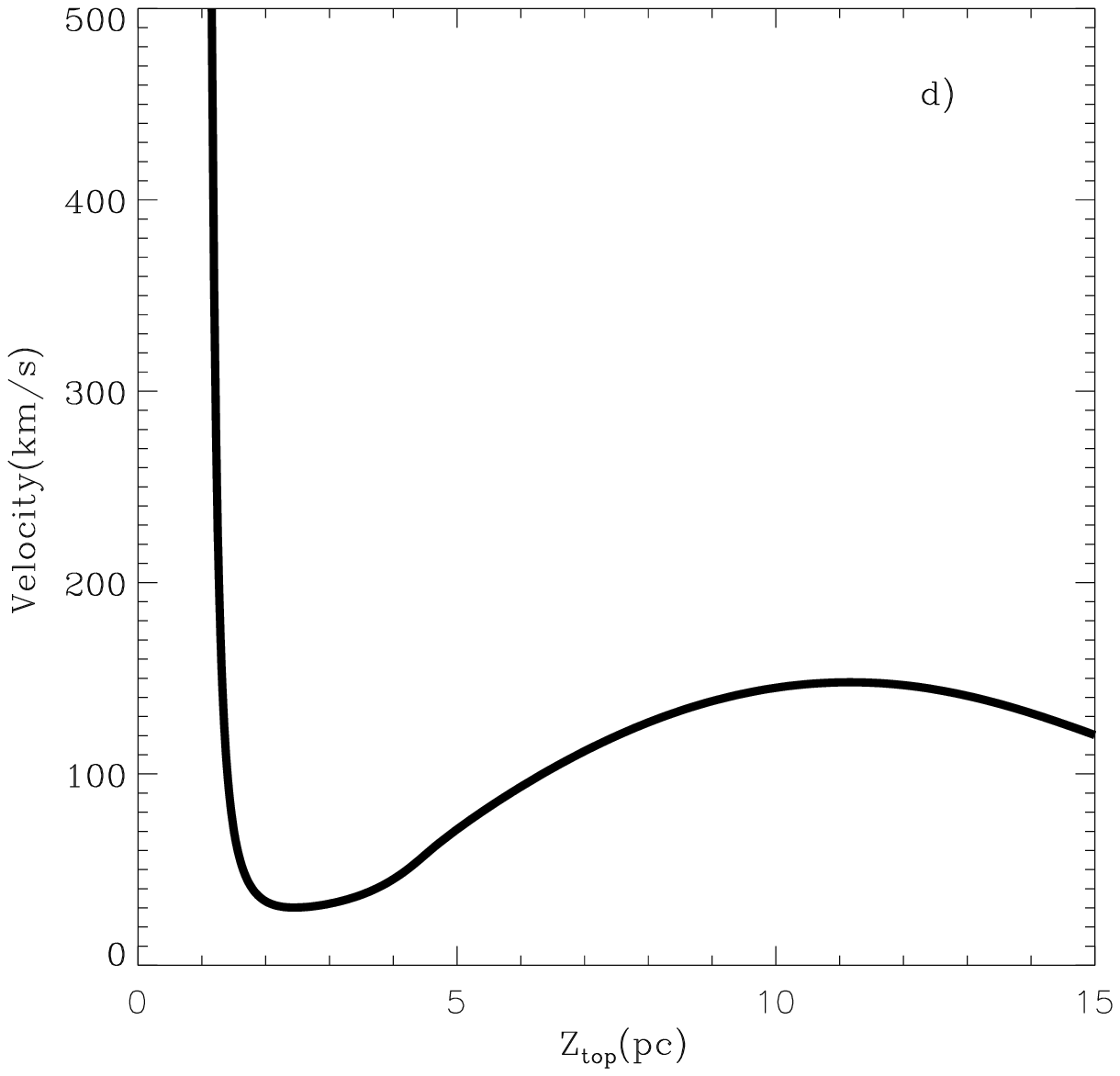}
\includegraphics{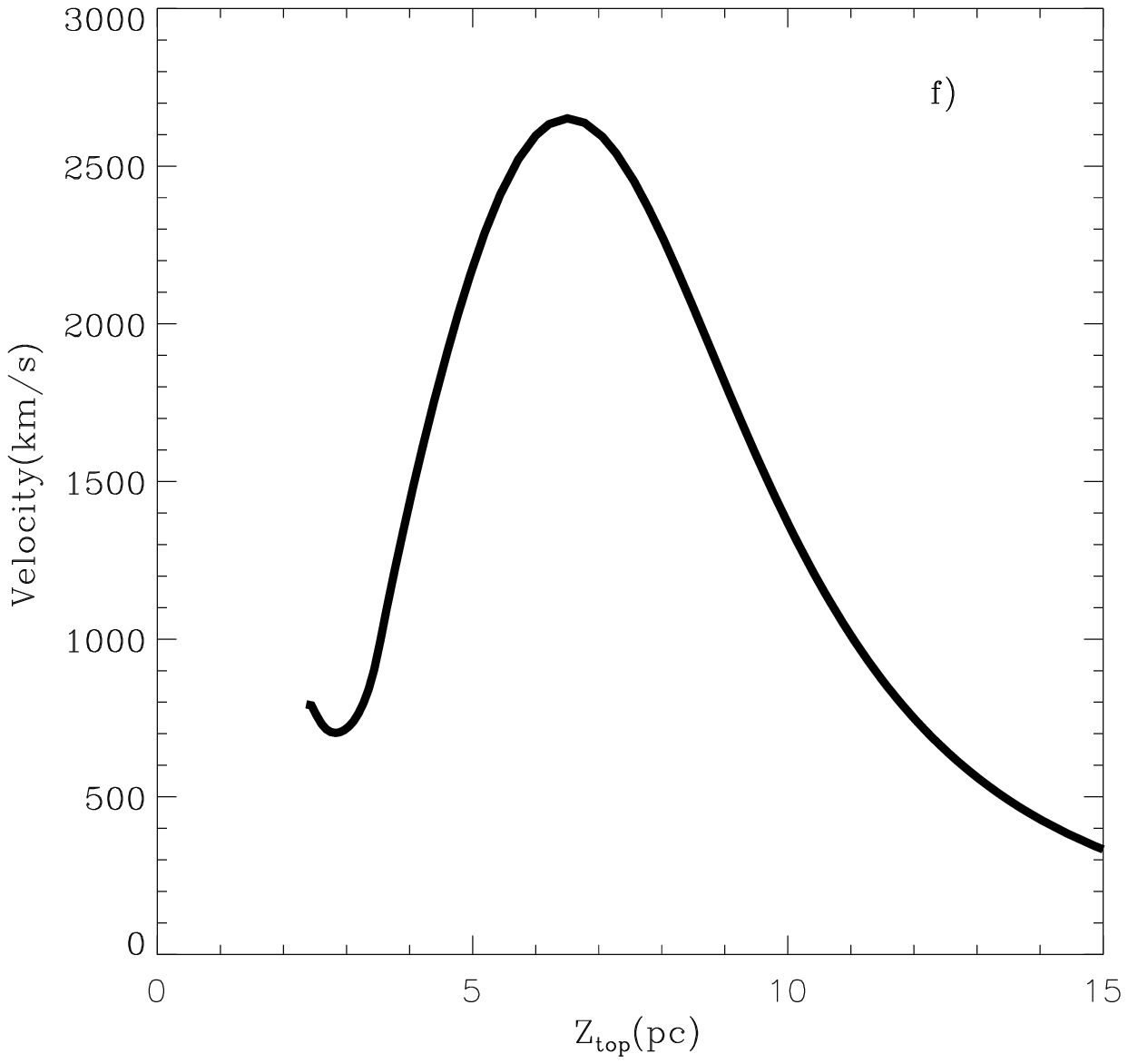}
\caption{The evolution of the volume occupied by the ejecta (left panesls) and the shock top pole velocity. Panels
         a and b, c and d, e and f show the results of the calculations when
         the explosion occurs at  $Z_{off} = 0$pc, 1pc and  2pc from the cluster center, respectively. The 
         solid, dotted and dashed lines in panel a correspond to 
        $10^4$yr, $5 \times 10^4$yr and $2 \times 10^5$yr after the 
         explosion. Panels c  displays
         the remnant shape at the times $5 \times 10^4$yr, 
         $10^5$yr and $1.5 \times 10^5$yr, and panel e at  $3 \times 10^3$yr, 
         $5 \times 10^3$yr and $7 \times 10^3$yr. Thin solid lines 
         display the gas density distribution in the remaining cloud, with the log of the density (cm$^{-3}$) indicated in every                      
         line.
         Panels b, d and f present the evolution of the shock top pole velocity.}
\label{f2}
\end{figure}

Note that as the SN blast wave sweeps the overtaken matter into a 
shell it also unveils a large fraction of massive stars from the 1SG. These 
are able to photoionize the inner edge of the surrounding shell, causing a 
mayor increase on its pressure. Its temperature would be $T_s \sim$ 10$^4$ K 
and its density is about $10^6$ cm$^{-3}$, while in the cavity interior the 
gas would present a temperature around
$10^6$ K and a density of the order of $10^{-2}$ cm$^{-3}$ and thus a strong 
inward champagne flow will take place 
\citep{Tenorio1979,Franco1990}.
The collapse velocity of the cavity is: $v_{collapse} = 
c_s (P_{shell}/P_{cavity})^{0.5}$, where $c_s$ is the sound speed in the 
photoionized shell ($\sim 10$ km s$^{-1}$) which leads to $v_{collapse}$ 
values  larger than 10$^3$ km 
s$^{-1}$.  The collapse would be even faster for cavities left after 
blowout by off-centered explosions as in all of these, the exit of the hot
gas will leave a neglishible pressure in the cavity allowing the photoionized 
shell to rapidly restore the original density. The collapse of the cavity is 
shorter 
than the spacing between SN, which for a $10^6$ M$_\odot$  cluster with a 
standard IMF occur about every 3000 years. This implies that sequential SN in 
clusters with a stellar mass $\le$ 10$^6$ M$_\odot$ will encounter a fully 
restored  gas density distribution.   

\section{Discussion}
\label{sec:3}

We have followed the consensus indicating that the peculiar light
element  anti-correlations,  the He enhancement ubiquitous to all GGC and 
particularly the Fe spread found in the most massive GC are signatures of 
2SGs resultant from matter hugely polluted by the ejecta
of a 1SG (see Renzini 2013, for a critical review on the 
various suggested 1SG polluters).

If one assumes a massive 1SG (M $\sim 10^5 - 10^6$ M$_\odot$) 
and a Salpeter  IMF, one expects during the first 40Myr of evolution a core 
collapse supernova every $\sim 10^4$ to a few 10$^3$ yr (Starburst 1999) and 
their impact has been expected to be devastating, disrupting  the mass left 
over from star formation. Here however we have shown that 
SNe exploding in a steep density gradient and  those taken place within the 
densest central regions but still being able to reach the cluster core radius 
($R_c$) with supersonic velocities, lead to blast waves able to undergo 
blowout.   
The volume affected by the explosions  becomes highly elongated as they 
progress into the gradient to 
reach the cluster edge within a few thousand to a few tens of thousand years 
and then vent their energy and their metals into the medium surrounding the 
cluster. Such events lead then  to no contamination of the gas left over from 
the formation of a 1SG. The rapid loss of pressure after discharging their 
metals 
into the surroundings, inhibits the dispersal of the left over cloud
leaving then no trace within the clusters of the SN events. Note also that 
blowout will also take place if the shock waves from other explosions find 
cavities similar to those shown in Figure 1, releasing their energy and metals
within a thousand years out of the cluster volume. We have shown that 
only well centered SN blast waves  evolving into a dense medium would attain 
subsonic velocities before reaching the cluster core radius and thus are 
likely to be retained by the cluster. Thus if a 2SG then forms, this should 
present an Fe spread as observed in the most massive GGC.  

We have also shown that the cavities generated by SN are likely to be refilled
 either after blowout or after the shell stalls, due to the large UV flux from
 all 
unveiled stars which through photoionization provide  the inner skin of the 
shell or cavity with a large pressure and
cause the  rapid restoration of the original density distribution. Thus 
sequential SN from massive clusters 
are likely to explode in very similar environments and the majority of them 
will experience blowout.

\section{Acknowledgments}

The authors thank the referee for valuable comments. This study was supported 
by CONACYT - M\'exico, grants 167169, and 131913  and by the Spanish Ministry 
of Science and Innovation  for the ESTALLIDOS collaboration  
(grants AYA2010-21887-C04-04 estallidos4 and 
AYA2013-47742-C4-2-P estallidos5). GTT also acknowledges the C\'atedra Severo 
Ochoa at the Instituto de Astrof\'isica de Canarias and 
the CONACYT grant 232876 for a Sabbatical leave. S C acknowleges financial 
support from PRIN-INAF 2014 (PI: S. Cassisi) and PRIN MIUR 2010-2011 (prot. 
2010LY5N2T) and the friendly hospitality at the IAC. The authors appreciate 
the science and discussions among the participants of  the ESTALLIDOS
Star Formation Feedback Workshop (IAC, Nov. 2014) which triggered a good 
number of ideas.

\bibliographystyle{apj}
\bibliography{gc1.bib}

\end{document}